# Effect of strain on the electrical transport and magnetization of the epitaxial $Pr_{0.5}Ca_{0.5}MnO_3/La_{0.5}Ca_{0.5}MnO_3/Pr_{0.5}Ca_{0.5}MnO_3$ trilayer structures.

P. Padhan, and W. Prellier

Laboratoire CRISMAT, CNRS UMR 6508, ENSICAEN,

6 Bd du Marehal Juin, F-14050 Caen Cedex, FRANCE.

## Abstract

Epitaxial trilayer structures consisting of two antifferomagnetic charge/orbital order insulators $Pr_{0.5}Ca_{0.5}MnO_3$ (PCMO) and $La_{0.5}Ca_{0.5}MnO_3$ (LCMO) are grown on (001)-oriented $SrTiO_3$ and $LaAlO_3$ substrates. In this trilayer series, a thin film of LCMO with various thicknesses is sandwiched between the two fixed thicknesses of PCMO. These samples show a Curie temperature with a hysteretic field dependent magnetization at 10 K, although the individual compounds are antiferromagnetic. The zero field electronic transport of all samples on LAO shows thermally activated behavior, while the thermally activated behavior is suppressed and a metal-like transport is appearing for the samples on STO as the LCMO layer thickness increases above 10 unit cell. We have discussed these magnetic and transport properties of the trilayer structures on STO and LAO by the interfacial effect due to the stabilized CO state and the vibration mode of Jahn-Teller distortion.

The physical properties in the multilayer structures based on the transition metal compounds are influenced by the structural and magnetic modification induced by the 3D-coordination of the transition metal ions, at the interfaces of the constituents [1, 2]. In these heterostructures, the interfaces are rich in magnetic and structural coordinations of the transition metal ions through the interaction processes like direct exchange, superexchange and double exchange. A manifestation of exchange coupling is the interfacial ferromagnetism at the interfaces of the heterostructures. Ueda et. al.[3] have studied the magnetic properties of the superlattices, consisting of antiferromagnetic(AFM) layers of $LaCrO_3$ and $LaFeO_3$ grown on (111)-oriented $SrTiO_3$ (STO), that show a ferromagnetic behavior. The authors have explained that the ferromagnetic behavior is due to the ferromagnetic coupling between $Fe^{3+}$ and $Cr^{3+}$. Takahashi at. al.[4] have also studied the transport and magnetic properties of the superlattices made up of AFM $CaMnO_3$ and paramagnetic $CaRuO_3$ grown on (001) oriented $LaAlO_3$(LAO). The resulting films show a Curie temperature ($T_C$) at ~ 95 K and a negative magnetoresistance below $T_C$. The authors have concluded that the ferromagnetic-like transition with appreciable spin canting occurs, only near the interface region, due to the electron transfer from the $CaRuO_3$ layer to the $CaMnO_3$ layer through the interface. These examples illustrate the importance of the interfaces in the properties of the oxide superlattices.

Mixed valance manganites exhibit several fascinating phenomena such as colossal magnetoresistance (CMR)[5], ordering of charge, orbital and spin of $Mn^{3+}$ and $Mn^{4+}$ [6,7,8] and electronic phase separation[9]. The charge ordering phenomena has been seen in particular, when the dopant concentration is close to the commensurate value x = 0.5 in the reduced bandwidth systems. In these systems, the charge ordering gap can be collapsed by the application of a magnetic field, an electric field, a high pressure, an optical radiation or an electron irradiation. This results in a metal-like transport below the charge order transition temperature. However, the interfaces of mixed valence manganites are more complex especially when two competing ground states, the FM metallic state and the charge ordered (CO)/orbital ordered (OO) insulating state, are present[9].

In our earlier studies, we have found the evidence of ferromagnetic phase and the enhanced coercive field in the superlattices consisting of two antiferromagnets $La_{0.5}Ca_{0.5}MnO_3$ (LCMO) and $Pr_{0.5}Ca_{0.5}MnO_3$ (PCMO) grown on (001) oriented $LaAlO_3$(LAO) [10]. In this article, we report the studies on the structure, magnetization and transport properties of the trilayer, where various thicknesses of LCMO layer sandwiched between two PCMO layers with different fixed thicknesses, simultaneously grown on (001) oriented LAO and $SrTiO_3$(STO).

The trilayer structures were grown on (001) oriented LAO and STO substrates using the multitarget pulsed laser deposition technique. The details of optimized deposition conditions are described elsewhere [10,11]. The deposition rates of PCMO and LCMO were calibrated for each laser pulse of energy density ~ 3 J/cm$^2$ from the superlattices reflection in x-ray $\theta$-$2\theta$ scans of the PCMO/LCMO superlattices. A series of trilayer structures were grown, where the bottom 50-(unit cell, u.c.) and top 10-u.c. thick PCMO layers sandwiched, n-u.c. of LCMO layer with n taking integer values from 1 to 18. To reduce the substrate induced strain, and, to have the effect of interface in the transport measurements, we have deposited a thicker bottom layer than the top layer. The epitaxial growth and the structural characterization of the trilayer structures were performed using x-ray diffraction. The magnetization (M) and magnetotransport measurements were performed using a superconducting quantum interference device based magnetometer (Quantum Design MPMS-5) and Physical property measurement system (Quantum Design PPMS) respectively.

The trilayer structures show (00*l*) Bragg's reflections of the constituents and the substrate, indicate the growth of epitaxial pseudocubic phase. The bulk pseudocubic lattice parameter of STO, LAO, PCMO and LCMO is 3.905 Å, 3.79 Å, 3.802 Å[6] and 3.83 Å[12] respectively. The STO provides an in-plane tensile stress for the epitaxial growth of PCMO with − 2.63 % lattice mismatch while the LAO provides an in-plane compressive stress with + 0.3 % lattice mismatch. These opposite substrate induced stress states on the bottom PCMO layer in the trilayer are seen from the relative fundamental peak positions of the sample and substrate in θ-2θ x-ray scan. In Fig.1, we show the θ-2θ x-ray scans close to the (001) and (103) reflections of the sample with n = 18 grown on (001) oriented STO (panel a and b respectively) and LAO (panel c and d). In order to study the in-plane epitaxy of these samples on STO and LAO, we have performed the φ-scans of the sample around asymmetric (103) planes. The reflection intensity from the asymmetric (103) planes of the trilayer samples with n = 18 grown on LAO and STO are shown in the Fig. 2a. The presence of four symmetric peaks at 90° interval confirms the four-fold symmetry of the pseudocubic perovskites. The negligibly small difference between the angular position of the peak in the φ-scans of the substrate and the film clearly shows the cube-on-cube growth morphology of the film. The out-of-plane lattice parameter of various samples extracted from the peak position in the θ-2θ x-ray scan is compared with the lattice parameter of PCMO and LCMO in the Fig. 2b. The lattice parameter of the sample is correlated with the interfacial and the substrate induced stress. The samples on LAO do not show a remarkable relaxation and variation of lattice parameter with the increase in LCMO layer thickness. While the lattice parameter of the samples on STO increases as the LCMO layer thickness increase from 3 u.c. and approaches to that of the bulk PCMO value for higher thickness.

The temperature dependent resistivity ρ(T) of bulk PCMO[6] and LCMO[12] show a thermally activated behavior while the temperature dependent magnetization M(T) shows paramagnetic-antiferromagnetic transition. These physical properties are remaining intact in the thin film deposited on STO and LAO[11, 13]. However, we have observed a change in their physical properties when a thin film of LCMO is sandwiched between two thin films of PCMO. The field-cooled (FC) temperature dependent magnetization after the diamagnetic correction to the substrate of four sandwiched structures with different LCMO layer thicknesses ($t_{LCMO}$) grown on STO and LAO, measured with 0.01 T at 10 K are shown in the Fig. 3. The sample with n = 5 on STO shows a paramagnetic-to-ferromagnetic transition at ~ 250 K and than a ferromagnetic-to-antiferromagnetic transition ($T_C$) at ~ 30 K while the sample simultaneously deposited on LAO shows $T_C$ at ~ 213 K. The corresponding values of $T_C$ seen for the samples on STO and LAO with n = 5 are remaining the same for the higher value of n. As the $t_{LCMO}$ increases above 5 u.c., the ferromagnetic-to-antiferromgnetic transition seen for the sample with n = 5 on STO, is suppressed and the magnetization at the ferromagnetic state increases. The magnetization of the ferromagnetic state of the samples on LAO increases as the value of n increases from 5 to 10 and shows a negligible change for the higher value of n. While the magnetization in the paramagnetic state of the samples on LAO increases for n > 10. This change in the magnetization in the ferromagnetic state of the sample on STO and LAO is consistent with the observed change in magnetization in their corresponding magnetic hysteresis loop (Fig. 4). In addition, The coercive field ($H_C$) of these samples on STO and LAO increases with the increase in n and saturate for n ≥ 10. The coercive field of the sample on LAO is smaller than the sample on STO. For example, the $H_C$ of the sample with n = 12 on STO and LAO are 0.033 T and 0.065 T respectively.

We have also studied the transport properties of these samples to understand the observed variation in their structural and magnetic properties. The $\rho(T)$ at 10 K in the presence of 0 T and 7 T magnetic field of the superlattices with n = 5, 10 and 18 deposited on STO and LAO are shown in the Fig. 5(a) and 5(b) respectively. On cooling below room temperature the zero-field ρ(T) of the sample with n = 5 on STO shows thermally activated behavior down to 100 K and at a temperature below 100 K, its resistance is limited by the input impedance of the PPMS. However, for the sample with n = 10 the resistivity below room temperature is insulator-like down to 10 K. While on cooling from room temperature the resistivity of the sample with n = 18 shows thermally activated behavior down to 150 K, shows metal-like behavior in the temperature range of 150 K to 30 K and an upturn below 30 K. In the presence of a 7 T magnetic field, the ρ(T) of the superlattice with n = 5 is similar to that of the zero-field ρ(T) of the sample with n = 18. However, as the n increases the metal-like window becomes broader with a higher metal-insulator transition temperature. The zero-field ρ(T) of all samples on LAO is qualitatively similar to that of the zero-field ρ(T) of the sample on STO with n = 5. The ρ(T) of these samples with n = 5 and 10 remains thermally activated in the entire temperature range even in presence of 7 T magnetic field. However, in presence of a magnetic field, on heating from 10 K the sample with n = 18 shows thermally activated behavior up to 120 K, on further heating a metal-like behavior appears in the temperature range of 120 K to 150 K and then remains insulator-like up to room temperature.

In order to understand the physical properties of these trilayer structures whose cumulative thickness is ≤ 308 Å, we first discuss the Jahn-Teller (JT) distortion of the $MnO_6$ octahedra. Using the in-plane lattice parameters of the trilayers extracted from the θ-2θ scans along the asymmetric (103) direction and the corresponding out-of-plane lattice parameter from the θ-2θ scans along the (00$l$) directions, we have calculated the lattice parameters of the orthorhombic structures the sample on STO and LAO. For the samples on STO the relation between the unit cell parameters is $c/\sqrt{2}$ (5.365 Å) < a ≈ b (5.42 Å) while for the sample on LAO it is $c/\sqrt{2}$ (5.44 Å) > a ≈ b (5.35 Å). This correlation of the unit cell parameters indicates the opposite nature of tetragonal distortion in the sample on STO and LAO, which results in an elongation or a contraction of the $MnO_6$ octahedron corresponding to the filled $3dz^2 - r^2$ orbital or $x^2 - y^2$ orbital, respectively. Since the orbital ordering, and hence the magnetic structures, is closely related to different types of cooperative JT vibration modes,[14] the apically compressed MnO$_6$ octahedra for the sample on STO imply a $3dx^2-r^2/3dy^2-r^2$ polarization of the $e_g$ orbitals, while the apically elongated octahedra for the sample on LAO imply a $3dz^2-r^2$ polarization of the $e_g$ orbitals [6,15]. So in the sample on STO and LAO the possible CO state is charge exchange (CE) type and C type respectively[16, 17].

The magnetization of the trilayer is due to the contributions from the bulk PCMO, LCMO layers and their interfaces. So, the key ingredients for the magnetic properties of the trilayer structures are the nature of stabilized CO – states and the interfacial spin configurations due to spin ordering[3], spin frustration[4] and spin canting[2]. The nature CO-state of the constituents may play an important role on the spin configurations i.e. the formation of the magnetic domain at the interfaces. The observed negligibly small changes in the low field moments at low temperature and the increase in the paramagnetic moments with LCMO layer thickness could be due to the C-type CO state of the constituents. The well defined Curie temperature and H$_C$ of the trilayer on LAO is due to the formation of the magnetic domain at the interfaces which is evidenced from ZFC/FC magnetic properties of the multilayer structure that have more interfaces[10]. The magnetization of the trilayer on STO and LAO increases gradually with the increase in magnetic field and does not show a clear saturation. The value of magnetization extracted from the hysteresis loop of the samples on STO and LAO with n = 18, taking into account of the weak diamagnetic response of the substrate, by extrapolating the linear part of the hysteresis loop to H = 0 are 1.981 $\mu_B$ and 0.399 $\mu_B$ respectively. This value of magnetization is very small compared to the theoretical value (3.5 $\mu_B$) of ferromagnetic phase of (Pr,La)$_{0.5}$Ca$_{0.5}$MnO$_3$ composition. To verify the stabilization of the ferromagnetic phases in PCMO and/or LCMO, we have measured the zero-field-cooled and field-cooled magnetic hysteresis loop (Fig. 4). The negligibly small change in the shape of the ZFC and FC hysteresis loop do not conclude the presence of ferromagnetic cluster as expected from the FM/AFM exchange bias system. However, the formation of ferromagnetic domains at the interfaces can not be ruled out[2,3]. The increase in the low field magnetization with the LCMO layer thickness, the higher value of magnetic moments and $H_C$ of the trilayer on STO compared to that of the simultaneously grown sample on LAO is associated with the possible CE-type CO state of the constituents. The stabilized CO state and the vibration mode of JT lattice distortion may partially be responsible for the strength of ferromagnetism and the phase separation tendency in the trilayer structures on STO[16].

The magnetic properties of the trilayer structures on STO and LAO are qualitatively similar but different from the constituent materials. However, the transport properties of the constituents are observed to be different in the trilayer simultaneously deposited on STO but not on LAO. The trilayer on STO shows metal-like behavior in the zero-field $\rho(T)$ as the LCMO layer thickness increases above 10 u.c. and the CO/OO state is partially lost even at 7 T magnetic field. While 7 T magnetic field is not enough to modify the CO/OO state to metal-like state of the sample with n < 18 on LAO. In PCMO and LCMO, the electronic transport is interpreted in terms of spin polarized tunneling conduction across the insulating phase and percolative conduction mechanism. Assuming that the transport in the PCMO/LCMO heterostructures is due to similar conduction processes, one can expect a similar magnetic and transport behavior in the heterostructures deposited on LAO and STO. The difference in the strength of CO/OO state in the trilayer on STO and LAO indicates that the interfaces and the substrate induced stress are the primary source of magnetism as well as the transport behavior of the heterostructures. The exchange coupling and transport are influenced by the crystallographic and/or magnetic reconstructions and relaxations due to the lattice mismatch and strain relaxation at the interfaces. This induced the possible vibration modes of JT distortion and the magnetic ordering may facilitate the occurrence of double exchange coupling and hence the metal-like transport of the sample on STO.

In conclusion, we have characterized and studied the transport and magnetic properties of the trilayer structures, where a thin film of LCMO is sandwiched between two thin films of PCMO grown on (001)-oriented LAO and STO. In the fixed PCMO layer based trilayer structure, the magnetic moment as well as the coercivity increase with the increase in LCMO layer thickness though the parent compounds are antiferromagnetic. A metal-like transport is appearing for the trilayer structures on STO as the LCMO layer thickness increases, though the parent compounds are insulator-like. While the electronic transport for the trilayer structure on LAO is thermally activated in the entire temperature range. We attribute this transport behavior to the possible vibration mode of JT lattice distortion due to the crystallographic and/or magnetic reconstructions and relaxations induced by the lattice mismatch and strain relaxation at the interfaces.

Acknowledgments:

We greatly acknowledged the financial support of the Centre Franco-Indien pour la Promotion de la Recherche Avancee/Indo-French Centre for the Promotion of Advance Research (CEFIPRA/IFCPAR) under Project No 2808-1.

Figure captions:

Figure 1: Typical room temperature θ - 2θ x-ray diffraction pattern around the (001) and (103) Bragg's diffraction peak of (50 u.c.) PCMO/(18 u.c.) LCMO/(10 u.c.) PCMO trilayer structure grown on (001) oriented STO (panel a and b respectively) LAO (panel c and d).

Figure 2: (a) Azimuthal dependence x-ray intensity recorded around the (103) angular position of the trilayer with n = 18 grown on (001) oriented STO and LAO. (b) The c-axis lattice parameter of the trilayer structures and the substrates. The solid lines are a guide to the eyes.

Figure 3: Field(0.1 T)-cooled temperature dependent magnetization of the (50 u.c.) PCMO/(n u.c.) LCMO/(10 u.c.) PCMO trilayer structures with n = 5, 10, 12 and 18 deposited on (001) oriented $SrTiO_3$ and $LaAlO_3$.

Figure 4: Zero-field-cooled, field dependent magnetization at 10 K of the (50 u.c.) PCMO/(n u.c.) LCMO/(10 u.c.) PCMO trilayer structures with n = 12 and 18 deposited on (001) oriented $SrTiO_3$ and $LaAlO_3$.

Figure 5: Zero-field-cooled temperature dependent resistivity in the presence of 0 T and 7 T magnetic field of (50 u.c.) PCMO/(n u.c.) LCMO/(10 u.c.) PCMO trilayer structures with n = 12 and 18 deposited on (001) oriented (a) $SrTiO_3$ and (b) $LaAlO_3$.

Fig. 1 : Padhan et. al.

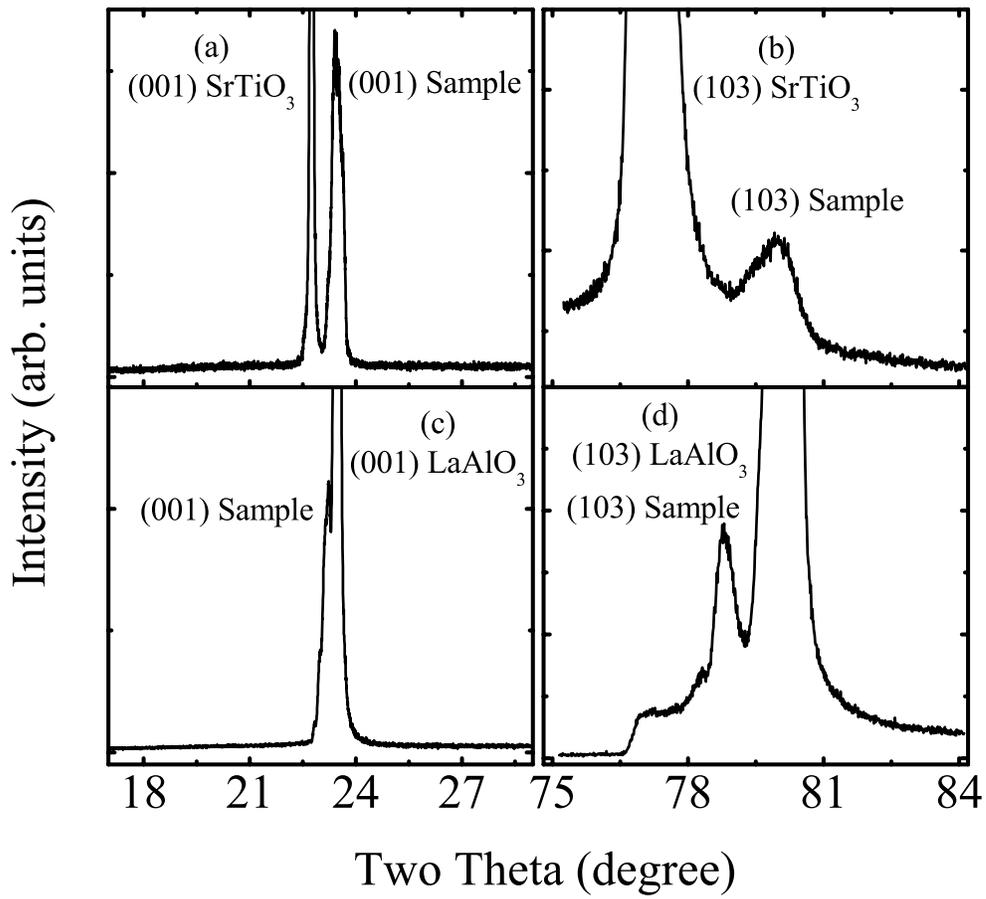

Fig. 2 : Padhan et. al.

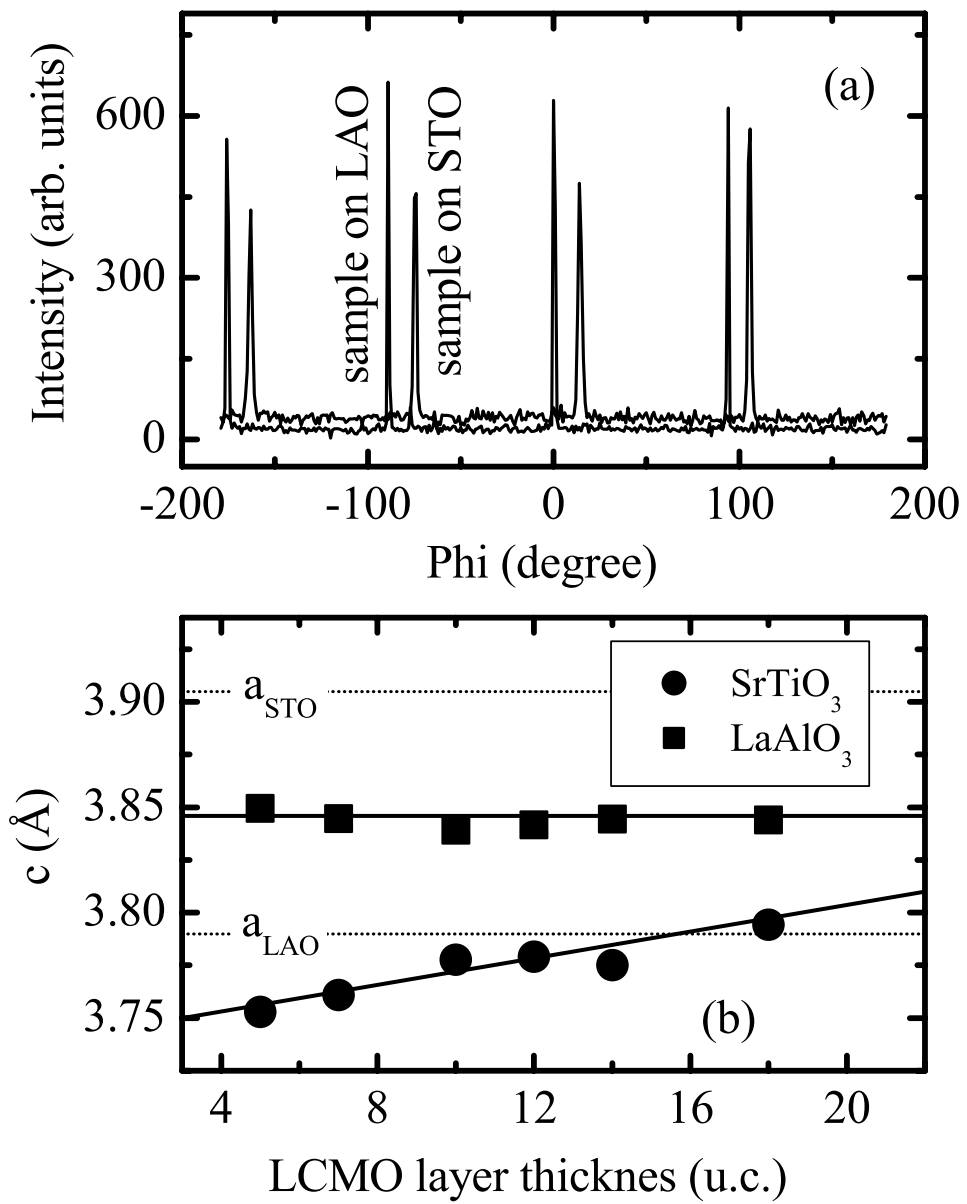

Fig. 3 : Padhan et. al.

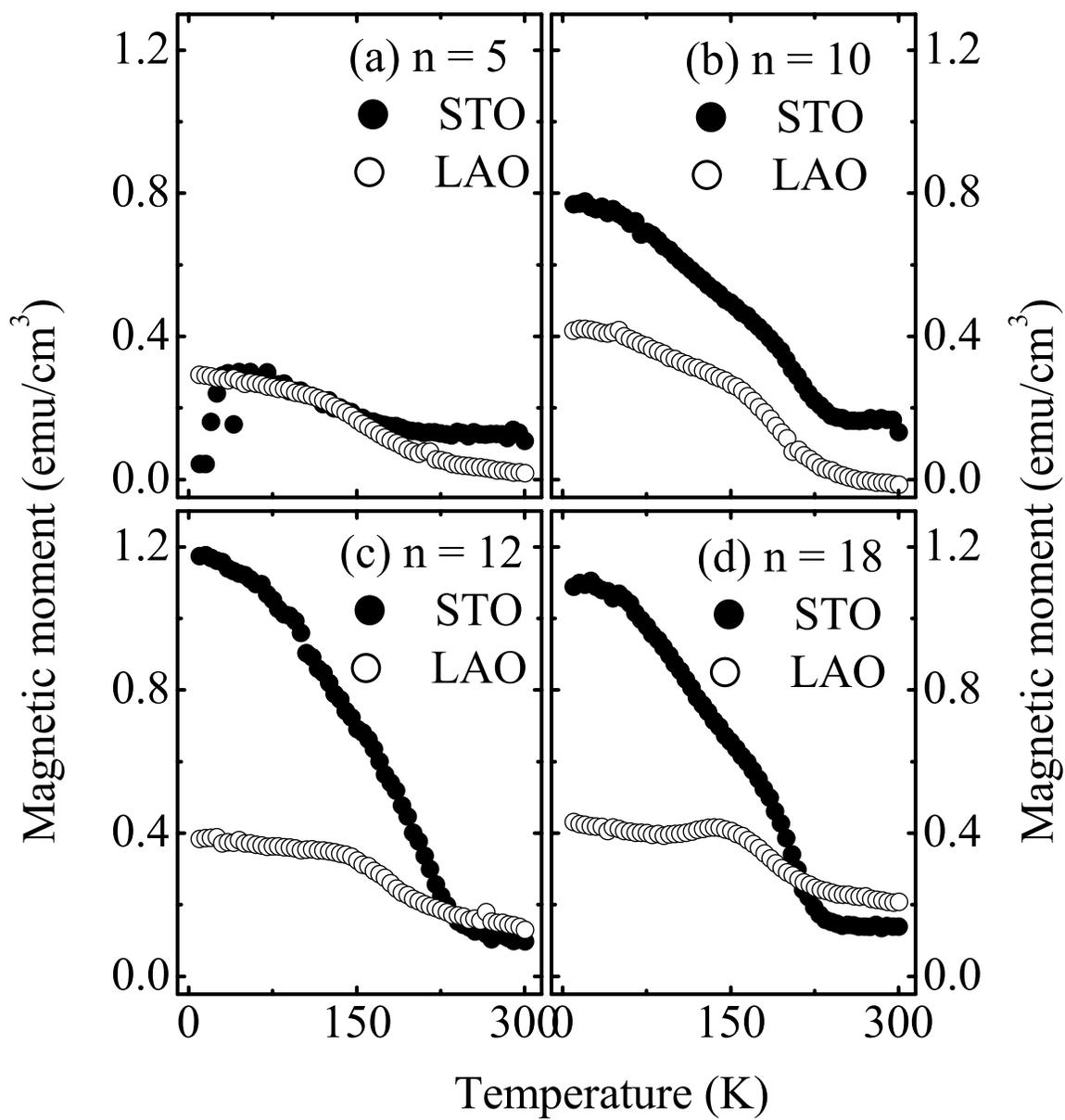

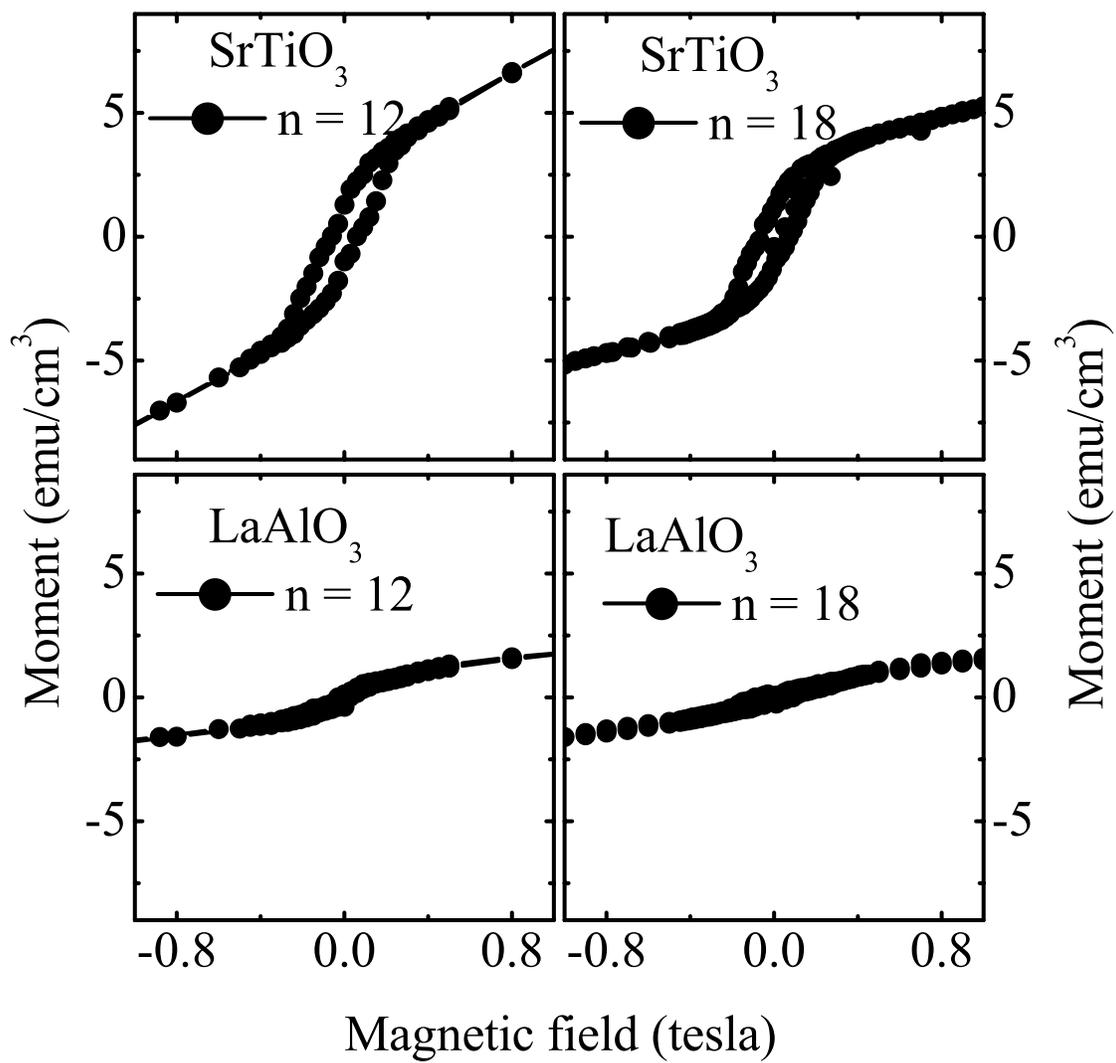

Fig. 5 : Padhan et. al.

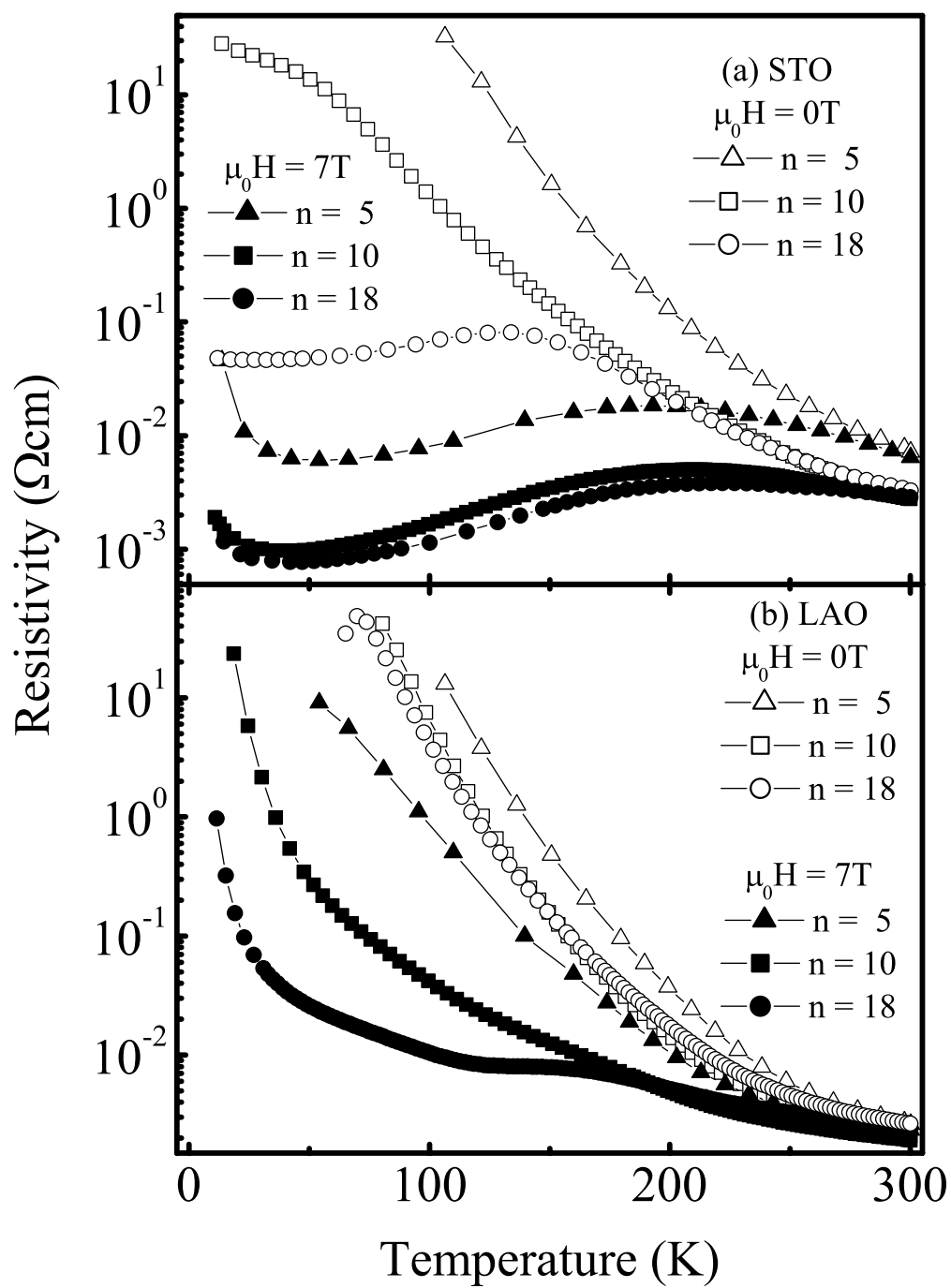